# Investigations of reactivity worth measurement in a fast neutron reactor with the inverse kinetics method[*]

WANG Shu-Miao(王姝妙)[1,2]   ZHOU Hao-Jun(周浩军)[1,2]   BAI Zhong-Xiong(白忠雄)[1,2]
FAN Xiao-Qiang(范晓强)[1,2]   YIN Yan-Peng(尹延朋)[1,2,1)]

1 Institute of Nuclear Physics and Chemistry, China Academy of Engineering Physics, Mianyang, Sichuan, China, 621900

2 Key Laboratory of Neutron Physics, China Academic of Engineering Physics, Mianyang, Sichuan, China, 621900

**Abstract:** A new method for the measurement of sample reactivity worth in a fast neutron reactor named the inverse kinetics method is proposed in the paper. The sample reactivity worth could be obtained by measuring the reactivity step change in the process of sample fetching and placing in the delayed critical reactor. Compared with the traditional period method, the advantage is that the accuracy of reactor reactivity control will not exert any influence on the uncertainty of reactivity worth measurement. The inverse kinetics method has been used to measure the reactivity worth of Φ20mm×9mm Au, V and Be samples at the center of the upper surface of a highly enriched uranium fast neutron reactor, and the results are 5.17 $\cent$, 4.40 $\cent$ and 5.90 $\cent$ respectively, which are consistent with those obtained with the period method. The standard uncertainty of measurement of the results is about 0.03 $\cent$, which achieves an obvious improvement compared with that (~0.08 $\cent$) of the period method.

## 1 Introduction

Compared with the effective neutron multiplication factor $k_{\text{eff}}$, the sample reactivity worth is more sensitive to the change of neutron nuclear reaction cross section, and the uncertainty of cross section has more significant influence on the uncertainty of the sample reactivity worth calculation in the sensitivity and uncertainty analysis. Therefore, measurement of the sample reactivity worth has always been an important macro integral experiment for inspection and adjustment of neutron nuclear reaction cross section.

The period method and the reactivity compensation method[1~3] are mainly used for the measurement of sample reactivity worth in a fast neutron reactor. These two methods have been brought into use since the 1950s and never changed. Our research group studied the sample reactivity worth measurement with the period method on a highly enriched uranium fast neutron reactor in 2013[4]. The reactivity worth of 13 kinds of metal samples such as Au, Fe and Ni at the center of the reactor upper surface was obtained with a standard uncertainty of measurement of 0.08 $\cent$. It is found that the main source of the measurement uncertainty of the period method is the reactor's reactivity control accuracy due to positions non-repeatability of the main drive and control rods of the reactor.

In order to further reduce the uncertainty of measurement of the sample reactivity worth, a new method named the inverse kinetics method is proposed in this paper which is based on the inverse kinetics reactivity measurement technology. It will realize fast and accurate measurement of the sample reactivity worth by measuring the reactor's reactivity step change in the process of online sample fetching and placing in the delayed critical reactor. The method was used to measure the reactivity worth of the samples such as Au on the upper surface of the highly enriched uranium reactor. The influence of the reactor's reactivity control accuracy on uncertainty is eliminated in the experiment and the combined standard uncertainty of measurement of the results is 0.03 ¢, which has been improved obviously compared to the traditional period method.

## 2 Inverse kinetics measurement system

The inverse kinetics method is a classical method for reactivity measurement. According to the point reactor kinetics equation, the inverse kinetics equation is written as Equation (1). Therefore, the change in reactivity over time can be obtained while the reactor's power $n(t)$ is known.

$$\rho(t) = \beta_{\text{eff}} + \frac{\Lambda}{n(t)} \frac{dn(t)}{dt} - \frac{\Lambda}{n(t)} \sum_{i=1}^{M} \lambda_i e^{-\lambda_i t} C_i(t_0) - \frac{\Lambda q}{n(t)} - \frac{1}{n(t)} \sum_{i=1}^{M} \lambda_i \int_{t_0}^{t} \frac{\beta_i}{\Lambda} e^{-\lambda_i(t-\tau)} n(\tau) d\tau \quad (1)$$

Here $n(t)$ is the neutron density, $\rho(t)$ is the reactivity, $\beta_{\text{eff}}$ is the effective delayed neutron fraction, $\beta_i$ is the delayed neutron fraction of the $i^{\text{th}}$ group, $\Lambda$ is the neutron generation time of the reactor, $C_i$ and $\lambda_i$ are the density and decay constants respectively of the precursor of the $i^{\text{th}}$ group of delayed neutrons, M is the group number of the delayed neutrons, $q$ is the outer neutron source intensity, and $t$ is the time.

The schematic of the inverse kinetics reactivity measurement system[5] is shown in Fig. 1. The γ compensation-type $^{10}$B ionization chambers are selected as the detectors, and the current signals are input to the industrial personal computer after being collected and converted by the programmable ammeters. The reactor's reactivity can be obtained from the equation (1) with the current and time data. The delayed neutron group parameters come from G. R. keepin's work[6]. The reactivity obtained according to Equation (1) is a relative value in $(1$=100 ¢).

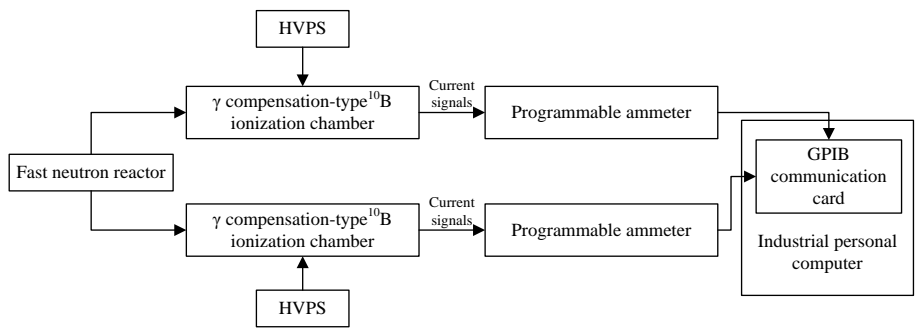

Fig. 1 Schematic structure of reactivity measurement system with the inverse kinetics method

The statistical variance $\sigma$ of reactivity measurement results with the inverse kinetics method is closely related to the current signal to noise ratio of the ionization chamber. The higher the neutron flux of the reactor, the smaller $\sigma$ will be. Consequently, the reactor should operate at a stabilized power in a delayed critical state when the inverse kinetics method is used to measure the sample reactivity worth.

## 3  Online sample fetching and placing system

The experiment was performed in the reactor's delayed critical state. Personnel shall not enter the reactor hall during the experiment. Therefore, an online automatic sample fetching and placing system was developed for sample operation. The system consisted of control cabinet, controller, rotary dividing plate and manipulator (including a chuck, a cylinder that can moving up and down and a cylinder that can moving front and back), as shown in Fig.2. The positioning accuracy of front, back, up and down transmission of the manipulator was 0.02mm. The rotary dividing plate with six storage positions was used for storage and automatic shift of samples and the tilt angle positioning accuracy was higher than 0.03°. The manipulator was installed next to the reactor and the chuck was 50cm from the reactor center. In order to realize automatic fetching and placing of sample between the sample sink of reactor and the rotary dividing plate, the rotary dividing plate and the upper surface of the reactor should be at the same level. Meanwhile, adjust the chuck, the rotary dividing plate's sample sink and the reactor's sample sink in a straight line by fine adjustment of the base position. The manipulator's control cable was laid out to the reactor control room, and a remote operation was done by a controller. A HD camera was installed near the reactor to make a real-time monitoring for the sample fetching and placing process in the control room during the experiment.

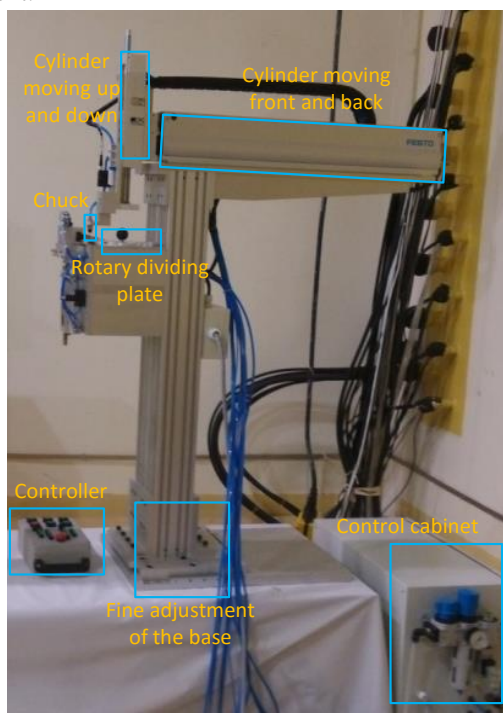

**Fig. 2 The automatic sample fetching and placing system**

## 4  Experimental process and results

The experiment was performed on a highly enriched uranium (HEU) fast neutron reactor. The core is composed of a control rod, a lower active zone, a middle steel disc, an upper active zone and reactivity adjustment components. There are four HEU reactivity adjustment components A, B, C and D at the top of the upper active zone. Another four components had been manufactured in the same dimensions but with stainless steel metal. The reactivity could be controlled by each adjustment component's presence and absence or changing its material. The center of the lower active section is a HEU metal control rod.

The experimental was conducted at the center of the upper surface of the reactor. The adjustment components A and B were removed and replaced by stainless steel components with a sample sink which was Φ20.00mm and 9.00mm deep. The samples, which were high purity Au, V and Be materials with the impurity content less than 0.01% provided by Beijing General Research Institute of Nonferrous Metals, were Φ19.98mm and 9.00mm thick.

In the experiment process, an Au sample was placed in the sample sink at first, and the reactor was operated to the delayed critical state with the power 30W and reactivity $\rho_{1,1}$ (thereafter, the reactivity under the same state should be denoted by $\rho_{1,i}$ in sequence, $i=1,2\cdots\cdots 10$). Afterwards, the Au sample was taken away. A positive step of reactivity with a very short time was recorded when the manipulator got near the sample and then the reactor changed to a subcritical state with the reactivity $\rho_{2,1}$ (thereafter, reactivity under the same state should be denoted by $\rho_{2,i}$ in sequence, $i=1,2\cdots\cdots 10$). Waited for 30s, the Au sample was put back into the sample sink again and the reactor returned to around the delayed critical state with the reactivity $\rho_{1,2}$, then waited for another 30s. Repeated the process above for 10 times, and the inverse kinetics measurement system was used to record the change of the reactivity over time, as shown in Fig. 3.

A local enlargement of the reactivity curve is presented in Fig. 3. The inverse kinetics curve would restore balance gradually for several seconds when the sample was just taking away or placing back into the sample sink. Therefore, only the last 20s within the 30s measurement time of each state can be used to calculate $\rho_{1,i}$ and $\rho_{2,i}$ by taking the average in such period.

It took about 23min to obtain the reactivity step curve shown in Fig. 3. The reactor temperature had increased 1.2℃ in the process. The sample reactivity worth can be calculated according to Equation (2) or Equation (3).

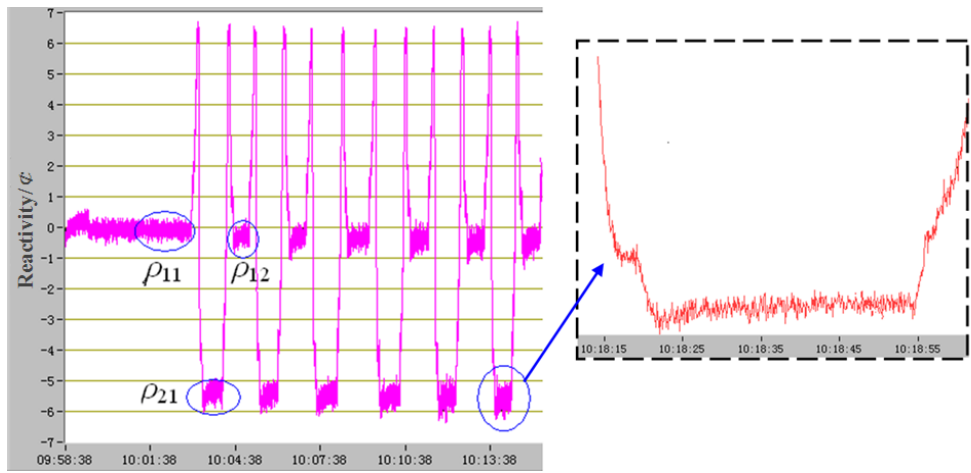

Fig. 3 Reactivity worth measurements of $^{197}$Au with the inverse kinetics method

$$\Delta\rho_i = \rho_{1,i} - \rho_{2,i} + k \cdot C \cdot \Delta t \qquad (2)$$

$$\Delta\rho_i = \rho_{1,i+1} - \rho_{2,i} - k \cdot C \cdot \Delta t \qquad (3)$$

Here $k$ is the average growth rate of reactor temperature in the experimental process, $\Delta t$ is the time interval for sample fetching and placing and $C$ is the temperature correction coefficient of the reactor, $C$=-0.38 ¢/℃.

The reactor temperature approximately increased at a constant speed in the experimental process. Therefore, the reactivity worth can be defined as follow by combining Equation (2) and Equation (3):

$$\Delta\rho_i = \frac{\rho_{1,i} + \rho_{1,i+1}}{2} - \rho_{2,i} \tag{4}$$

The Au sample's $\rho_{1,i}$, $\rho_{2,i}$ and the reactivity worth $\Delta\rho_i$ calculated according to equation (4) is listed in table 1. $\rho_{1,i}$, $\rho_{2,i}$ and $\Delta\rho_i$ keep four decimal places to keep more statistics information of experimental data. The Au sample reactivity worth is 5.17 ¢, obtained by taking the arithmetic mean value of $\Delta\rho_i$, with a Bessel standard deviation of 0.01 ¢. Same method is used to process other experimental data, and all results of the reactivity worth experiment are listed in table 2.

Table 1 Measurements of $^{197}$Au with the inverse kinetics method

| $i$ | 1 | 2 | 3 | 4 | 5 | 6 | 7 | 8 | 9 | 10 | 11 |
|---|---|---|---|---|---|---|---|---|---|---|---|
| $\rho_{1,i}$/¢ | -0.2246 | -0.2604 | -0.2891 | -0.3332 | -0.3716 | -0.4107 | -0.4582 | -0.5076 | -0.5504 | -0.6168 | -0.6602 |
| $\rho_{2,i}$/¢ | -5.4191 | -5.4513 | -5.4887 | -5.5210 | -5.5635 | -5.6017 | -5.6505 | -5.7062 | -5.7526 | -5.8029 | --- |
| $\Delta\rho_i$/¢ | 5.1766 | 5.1766 | 5.1775 | 5.1686 | 5.1723 | 5.1672 | 5.1675 | 5.1772 | 5.1690 | 5.1644 | --- |

Table 2 Measurements of the reactivity worth with the inverse kinetics method

| Samples | Au | V | Be |
|---|---|---|---|
| Reactivity worth/¢ | 5.17 | 4.40 | 5.90 |
| Bessel standard deviation/¢ | 0.01 | 0.02 | 0.01 |

## 5  Uncertainty of measurement

In the experimental process of the inverse kinetics method, the positions of main transmission and control rods of the reactor were kept fixed. Therefore, the influence of reactivity control accuracy on the measurements can be neglected. In the sample fetching and placing process, the difference between sample radius and sample sink radius is very small (<0.02mm). The influence of sample position non-repeatability in the sink is evaluated by considering the sample size, neutron flux distribution of experimental position and reactivity worth measurement results, and the resulting uncertainty for the experimental result is less than 0.001 ¢ which can be omitted. The influence of the uncertainty of $\Lambda/\beta_{\text{eff}}$ in equation (1) on the measurements can be omitted for that the item including $\Lambda/\beta_{\text{eff}}$ is five orders of magnitude less than others. In addition, the influence of temperature can also be omitted for the temperature correction item has been eliminated in Equation (4) during data processing.

Consequently, the uncertainty of measurement of sample reactivity worth by the inverse kinetics method is mainly from two aspects: the uncertainty of statistics of experimental measurement and the uncertainty of the delayed neutron group parameters.

The experimental result of sample reactivity worth listed in table 2 is obtained by taking mean of multiple measurements. Therefore, the statistic uncertainty of measurements can be expressed with the Bessel standard deviation divided by $\sqrt{n}$ where $n$ is the number of measurement times.

To evaluate the influence of the uncertainty of delayed neutron group parameters on the measurement result, the covariance of group parameters is necessary. Only the experimental measurement error[6] can be used for analysis in this paper because the covariance of the G. R. keepin's delayed neutron group parameters cannot be obtained. Let the six-group delayed neutron group parameters vary at will within the error range, and apply the group parameters to the reactivity calculation code with the inverse kinetics method. The result shows that the uncertainty of reactivity worth measurements caused by the uncertainties of the group parameters is no more than 0.03 ¢.

As a consequence, the combined standard uncertainty of measurement $u_c$ of sample reactivity worth in the experiment is

$$u_c = \sqrt{\frac{0.02^2}{10} + 0.03^2} = 0.031 ¢ \tag{5}$$

## 6  Experimental comparison

The reactivity worth measurement results at the center of upper surface of reactor coincide well with the result of the period method in 2013 for the same samples, and the difference between the measured values is less than 0.5%.

The uncertainty of the sample reactivity worth measurement of this experiment, the experiment with the period method in 2013 and the experiment carried out by CEA is listed in Table 3. CEA used CALIBAN and SILENE reactors to conduct the sample reactivity worth experiment from 2008 to 2012. The CALIBAN consists of four control rods with the reactivity control accuracy of each rod about 0.015 ¢. Two control rods were used in the experiment, and the combined control accuracy is about 0.02 ¢. SILENE is a solution reactor with a lower reactivity control accuracy of about 0.1 ¢. The HEU reactor used in this experiment has relatively lower reactivity control accuracy compared with CALIBAN. Only under specific operation conditions, can the reactivity control accuracy reach 0.05 ¢. Generally, it is about 0.06 ¢. The sample reactivity worth experiment of CEA didn't consider the influence of the delayed neutron group parameters. Actually the influence of the uncertainties of the delayed neutron group parameters on the uncertainty of measurement increases with the increase of the measurements when the period method is used. The sample reactivity worth measurements on the CALIBAN exceeds 5 ¢ and the uncertainty of measurement caused by the delayed neutron group parameters is about 0.06 ¢. Therefore, considering the influence of the delayed neutron group parameters, the combined standard uncertainty of the sample reactivity worth measurement with the period method has little difference between the HEU reactor and the CALIBAN.

Compared with the period method, the inverse kinetics method achieves a great improvement in the uncertainty of measurement of sample reactivity worth. Not only the measurement process would not be influenced by the reactivity control accuracy, but also the influence of the uncertainties of the delayed neutron group parameters on the uncertainty of measurement is reduced. The reactivity worth of Au and Be listed in table 2 also exceed 5 ¢, but the uncertainty of measurement caused by the uncertainties of the delayed neutron group parameters is only 0.03 ¢.

Table 3 Comparison of the uncertainty of measurement of the reactivity worth

| Experiment facility | Method | Main source of the uncertainty | Uncertainty of measurement/¢ | |
|---|---|---|---|---|
| | | | The delayed neutron group parameters not considered | The delayed neutron group parameters considered |
| HEU reactor | Inverse kinetics method | Uncertainty of the delayed neutron group parameters | ~0.01 | 0.031 |
| HEU reactor | Period method | Reactivity control accuracy | 0.05 | 0.08 |
| CALIBAN | Period method | Reactivity control accuracy | 0.02 | --- |
| SILENE | Period method | Reactivity control accuracy | 0.1 | --- |

## 7  Conclusions

The reactivity worth experiment is an important macro integral experiment used for inspection and adjustment of neutron nuclear reaction cross section. In this paper, a new method for the measurement of sample reactivity worth is proposed, namely the inverse kinetics method. The sample is placed somewhere in the reactor. A manipulator is used to fetch and place the sample automatically in the reactor's delayed critical state. The sample reactivity worth is obtained by using the inverse kinetics method to measure the reactivity step change. Compared with the traditional period measurement method, the inverse kinetics method eliminates the influence of reactivity control accuracy on the uncertainty of measurement of sample reactivity worth and the influence caused by the uncertainties of the delayed neutron group parameters is also reduced during data processing.

The reactivity worth of Φ20mm×9mm Au, V and Be samples at the center of upper surface of the HEU fast neutron reactor have been measured with the inverse kinetics method, which are 5.17¢, 4.40¢ and 5.90¢ respectively, and the combined standard uncertainty is about 0.03¢. The results coincident with those obtained with the positive period method under the same conditions and have an obvious improvement in the uncertainty of measurement.